\documentclass[11pt]{article}
\usepackage{graphicx}
\usepackage{epsfig}
\usepackage{array}
\usepackage[figuresright]{rotating}
\usepackage{amsmath}
\usepackage{amssymb}
\usepackage{amsfonts}

\newcommand{\veceps}{\mbox{\boldmath$\epsilon$}}
\newcommand{\vecsig}{\mbox{\boldmath$\sigma$}}

\usepackage{cite}
\begin{document}
\begin{center}
{\LARGE{\bf Structure of the nucleon and spin-polarizabilities} 
  }\\[1ex] 
Martin Schumacher\footnote{Electronic address: mschuma3@gwdg.de}\\
Zweites Physikalisches Institut der Universit\"at G\"ottingen,
Friedrich-Hund-Platz 1\\ D-37077 G\"ottingen, Germany\\
M.I. Levchuk\footnote{Electronic address: levchuk@dragon.bas-net.by}\\
B.I. Stepanov Institute of Physics, 220072 Minsk, Belarus
\end{center}
\begin{abstract}
Spin-polarizabilities are predicted by calculating the cross-section
difference $\sigma_{3/2}-\sigma_{1/2}$ from available data for the resonance
couplings $A_{3/2}$ and $A_{1/2}$ and CGLN amplitudes. The forward
spin-polarizabilities are predicted to be $\gamma^{(p)}_0=-0.58\pm 0.20$ and
$\gamma^{(n)}_0=+0.38\pm 0.22$ in units of  $10^{-4}$ fm$^4$
where the different signs  are  found to be due to the 
isospin dependencies of the $E_{0+}$ and the $(M,E)^{(1/2)}_{1+}$ amplitudes. 
The backward
spin-polarizabilities are predicted to be   $\gamma^{(p)}_\pi=-36.6$
and  $\gamma^{(n)}_\pi=+58.3$, to be compared with the experimental values  
 $\gamma^{(p)}_\pi=-36.4\pm 1.5$ and 
 $\gamma^{(n)}_\pi=+58.6\pm 4.0$. Electric $\gamma_E$ and magnetic 
$\gamma_M$ spin-polarizabilities are introduced and discussed in terms of the  
$E1$ and $M1/E2$ components  of the photo-absorption cross section of the nucleon.
\end{abstract}

\section{Introduction}
Studies  of the spin-polarizabilities of the nucleon are a fascinating
subject of present and future experimental and theoretical research.
In a recent article \cite{pasquini10} the forward amplitude for polarized
Compton scattering by dispersion integrals has been constructed as a guideline
for future experiments to extract the  spin-polarizabilities of
the nucleon. In this work \cite{pasquini10}
the spin-polarizabilities and their generalized
versions are given in terms of pion photoproduction multipoles 
as well as measured spin-dependent cross sections. Furthermore, 
references are  given of previous experimental and theoretical work.

The purpose of the present paper is to supplement on the important topic of
research addressed in \cite{pasquini10} by making use of methods to describe
the electromagnetic structure of the nucleon as derived in 
\cite{schumacher05,schumacher06,schumacher07a,schumacher07b,schumacher08,schumacher09,schumacher10}.
This method describes the amplitudes for Compton scattering in terms of
partial contributions which each can be attributed to a special one-photon
or two-photon excitation mechanism. The method is described in detail
in \cite{schumacher09} and also some more
information is given in the next section. The progress achieved in comparison
to previous approaches is based on an isospin decomposition of the CGLN
amplitudes described in \cite{schumacher09}. This makes it possible to avoid
the use of available data in terms of partial reaction channels. Through
this method it is possible to take advantage of the very precise
parametrizations of CGLN amplitudes of Drechsel et al.
\cite{drechsel07} which are  based on the world data of pion photo- and 
electroproduction and is updated regularly.
Furthermore, a method is developed to calculate total cross sections
$\sigma_T(\omega)$ and spin-dependent cross-section differences 
$\sigma_{3/2}(\omega)-\sigma_{1/2}(\omega)$ for the nucleon
resonances from the  resonance couplings $A_{3/2}$ and $A_{1/2}$ which are
collected and updated regularly by the Particle Data Group \cite{PDG}.
These resonance couplings contain the effects of the two-pion channel.
Therefore, the  use of these resonance couplings avoids the problem of
the two-pion channel which occurs when  relying on the CGLN amplitudes alone.
On the basis of these data it is possible to calculate any nucleon
structure quantity which is given by a dispersion integral not only
in terms of  the total quantity but separately in terms of multipolarity 
and isospin and in terms of  resonant or nonresonant excitations. This means
that the method translates one-to-one the world data \cite{drechsel07}
known from 
photoexcitation experiments into structure constants like the contributions
to the Gerasimov-Drell-Hearn (GDH) integral and the electric and magnetic 
polarizabilities and the spin-polarizabilities for the forward and backward
directions.

In the previous paper \cite{schumacher09} this method was applied to those
quantities where independent directly measured experimental data are
available. In a first place these are the cross-section differences 
$\sigma_{3/2}(\omega)-\sigma_{1/2}(\omega)$ measured by the GDH Collaboration
at MAMI (Mainz) and ELSA (Bonn). Good agreement has been obtained
between experimental data and predictions. One remarkable result is
that the cross-section difference $\sigma_{3/2}(\omega)-\sigma_{1/2}(\omega)$
measured for the $P_{33}(1232)$ resonance is strongly enhanced by the
$E2/M1$ ratio. This  enhancement amounts to $26\%$   whereas the corresponding
quantity is small for  $\sigma_T(\omega)$. This interesting finding
will be analyzed in more detail in the present paper. Another result obtained
in the previous paper \cite{schumacher09} was that the predicted GDH integrals
for the proton and the neutron are in good  agreement with the 
experimental results obtained at MAMI and ELSA. An investigation has also been
carried out for the backward spin-polarizabilities where again  good
agreement has been obtained with directly measured experimental data.
However, no investigation has been carried out for the forward
spin-polarizabilities because no directly measured data are available.
It is the purpose of the present paper to fill this gap of information
as far as the predictions are concerned.
Since again this investigation of the forward spin-polarizabilities
is based on the world data of CGLN amplitudes and resonance couplings
$A_{3/2}$ and $A_{1/2}$ the results are expected to be of good
precision.

\section{R\'esum\'e of dispersion theory }

\begin{figure}[h]
\centering\includegraphics[width=0.6\linewidth]{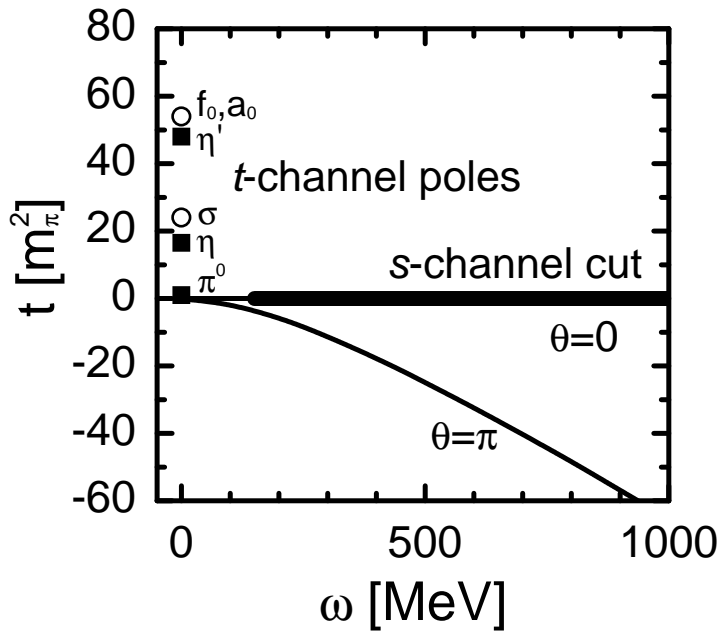}
\caption{Singularities  of the $s$- and the $t$-channel. Horizontal thick line:
$s$-channel cut representing the nucleon degrees of freedom, with $\omega$
being the energy of the incoming photon in the lab system. Squares and circles
on the 
vertical real $t$-axis: $t$-channel poles representing the constituent-quark
degrees of freedom. The physical range of nucleon Compton scattering extends
from the line at $\theta=0$ to the line at $\theta=\pi$.}
\label{s-t-plane}
\end{figure}

Compton scattering is  described by the invariant amplitudes $A_i(s,t)$
$(i=1-6)$ which
are analytic functions in the two variables $s$ and $t$  and, therefore,
may be treated in terms of dispersion relations 
\cite{lvov97,drechsel03,lvov99,hearn62}. 
The degrees of freedom of the nucleon including the structure 
of the constituent quarks enter into these  invariant
amplitudes via a cut on the real axis of the complex  $s$-plane and in terms
of point like 
 singularities on the positive real axis of the 
$t$-plane as illustrated in Figure \ref{s-t-plane}. 
The $s$-channel cut  contains
the total photo-absorption cross section and through a decomposition in terms
of excitation mechanisms the complete electromagnetic 
structure of the nucleon as seen in one-photon processes.
The point like singularities correspond to the bare masses 
$m_0$ of the mesons which in case of scalar mesons 
have to be compared with the masses entering into the pole
$\sqrt{s_R}=M_R-\frac12\,i\,\Gamma_R$ located on the second Riemann sheet
of the complex $s$-plane. For the $\sigma$-meson the bare mass is
$m_\sigma=666$ MeV \cite{schumacher10} whereas for the quantities
$M_\sigma$ and $\Gamma_{\sigma}$ the results $M_\sigma=441^{+16}_{-8}$ MeV
and $\Gamma_\sigma=544^{+18}_{-25}$ \cite{caprini06} have been obtained.
Further details may be found in \cite{schumacher10}.
The relation  between the bare mass $m_0$ and the pole at $\sqrt{s_R}$
can easily  be understood from the arguments contained in 
\cite{tornqvist95,boglione02}. Here the propagator is written down in the form
\begin{equation}
P(s)=\frac{1}{m^2_0+\Pi(s)-s}
\label{propagator}
\end{equation}
where $m_0$ is the bare mass of the scalar meson and $\Pi(s)$ a complex 
analytic function
taking into account the effects of decay of the scalar meson into two
mesons. The real part of $\Pi(s)$ is equal to the square of the mass shift
and the imaginary part proportional to the particle width. The analytic 
continuation of this 
propagator
has a pole at $s_R=(M_R-\frac12\,i\,\Gamma_R)^2$ on the second Riemann sheet. 
In case of Compton
scattering the reaction $\gamma\gamma\to \sigma \to N\bar{N}$ has to be 
considered 
instead of the reaction $\gamma\gamma\to\sigma \to \pi\pi$. This leads to the
consequence that $\Pi(s)\equiv 0$, so that the propagator has the form
\begin{equation}
P(s)=\frac{1}{m^2_0 -s},
\label{propagator1}
\end{equation}
corresponding to a pole on the positive $t$-axis at $t_0=m^2_0$.
Differing from the present procedure,
in older works (see \cite{schumacher05}) use is made of the pole 
located on the second Riemann sheet. This is possible by taking into account
the reaction
$
\gamma\gamma\to \sigma \to \pi\pi \to \sigma \to N\bar{N}
$
instead of the reaction $\gamma\gamma\to \sigma \to N\bar{N}$.
This older procedure is equivalent in principle  to the present one
but requires a more elaborate 
numerical calculation.

 The 
point like singularities on the positive real $t$-axis may be related 
to the structure of the constituent quarks which couple to all mesons
with a nonzero meson-quark coupling constant. Of these the $\pi^0$- and the
$\sigma$-meson are of special interest but also the mesons
$\eta$, $\eta'$, $f_0(980)$ and $a_0(980)$ have to be taken into account.
In a formal sense the singularities on the positive real $t$-axis
correspond to the fusion of two photons with 4-momenta $k_1$ and $k_2$
and helicities $\lambda_1$ and $\lambda_2$ to form a $t$-channel intermediate
state $|\pi^0\rangle$ or $|\sigma\rangle$ or one of the other mesons,
from which -- in a second step --
a proton-antiproton pair is created. In the present case the
$N\bar{N}$ pair creation-process is virtual, i.e. the energy is too low to put
the  proton-antiproton pair on the mass shell (see \cite{schumacher05} for
more details).

\subsection{The kinematics of Compton scattering \label{2.5}}

In more quantitative terms the content of the forgoing paragraph
may be described in the
following form.
The conservation of energy and momentum in nucleon Compton scattering
\begin{equation}
\gamma(k,\lambda) + N(p) \to \gamma'(k',\lambda')+N'(p')
\label{s-channel}
\end{equation} 
 is given by 
\begin{equation}
k+p=k'+p'\label{enermomen}
\end{equation}
where $k$ and $k'$ are the 4-momenta of the incoming and outgoing photon
and $p$ and $p'$ the 4-momenta of incoming and outgoing proton.
Mandelstam variables are introduced via
\begin{eqnarray}
&&s=(k+p)^2=(k'+p')^2, 
t=(k-k')^2=(p'-p)^2, 
u=(k-p')^2=(k'-p)^2,\\
&&s+t+u=2m^2. \label{constraint}
\end{eqnarray}
with $m$ being the nucleon mass.

In terms of the Mandelstam variables the 
scattering angle $\theta$ in the c.m.
system is given by
\begin{equation}
\sin^2\frac{\theta}{2}=-\frac{st}{(s-m^2)^2}.
\label{scatteringangle}
\end{equation}
The $t$-channel
corresponds to the fusion of two photons with four-momenta
$k_1$ and $k_2$ and helicities $\lambda_1$ and $\lambda_2$ to form
a $t$-channel intermediate state $|t\rangle$ from which -- in  a
second step -- a proton-antiproton pair is created.  
The corresponding reaction may be formulated in the 
form
\begin{equation}
\gamma(k_1,\lambda_1) + \gamma(k_2,\lambda_2)  \to  \bar{N}(p_1)  +N(p_2).
\label{t-channel}
\end{equation}
Since for Compton scattering
the related $N\bar{N}$ pair creation-process is  virtual,
in dispersion theory we have to treat the process described in
(\ref{t-channel})  in the unphysical region. 
In the c.m. frame of (\ref{t-channel}) where
\begin{equation} 
{\bf k_1}+{\bf k_2}=0 \label{k1plusk2}
\end{equation}
we obtain
\begin{equation}
\sqrt{t}=\sqrt{(k_1 + k_2)^2}= \omega_1 + \omega_2 = W^t,
\label{t-interpretation}
 \end{equation}
where  $W^t$ is the energy transferred to the $t$-channel via two-photon
fusion. At positive $t$ the $t$-channel
of Compton scattering $\gamma N\to N \gamma$ coincides with the $s$-channel
of the two-photon fusion reaction $\gamma_1+\gamma_2 \to N\bar{N}$.

\subsection{Dispersion integrals for the polarizabilities}

 For the
following discussion it is convenient to use the lab frame and to
consider special cases for the scattering amplitude $T_{fi}$.
These special cases are
the extreme forward ($\theta=0$) and extreme backward ($\theta=\pi$)
direction  where the amplitudes for Compton scattering
may be written in the form
\cite{babusci98}  
\begin{eqnarray}
&&\frac{1}{8\pi m}[T_{fi}]_{\theta=0}
=f_0(\omega){\veceps}'{}^*\cdot{\veceps}+
g_0(\omega)\, \mbox{i}\, {\vecsig}\cdot({\veceps}'{}^*\times
{\veceps}), \label{T0}\\
&&\frac{1}{8\pi m}[T_{fi}]_{\theta=\pi}
=f_\pi(\omega){\veceps}'{}^*\cdot{\veceps}+
g_\pi(\omega)\, \mbox{i}\,{\vecsig}\cdot({{\veceps}}'{}^*
\times
{{\veceps}})
\label{Tpi}
\end{eqnarray}
with $m$ being the nucleon mass, ${\veceps}$ the polarization of the photon
and $\vecsig$ the spin vector.
Equations (\ref{T0})
and (\ref{Tpi}) can be used to define the 
polarizabilities and spin-polarizabilities as the lowest-order
coefficients in an $\omega$-dependent development of the
nucleon-structure dependent parts of the scattering amplitudes:
 \begin{eqnarray}
f_0(\omega) & = & - ({e^2}/{4 \pi m})Z^2 + 
{\omega}^2 ({\alpha}
+{\beta}) + {\cal O}({\omega}^4), \label{f0}\\
g_0( \omega) &=&  \omega\left[ - ({e^2}/{8 \pi m^2})\,   
{\kappa}^2 
+ {\omega}^2
{\gamma_0}  + {\cal O}({\omega}^4) \right], \label{g0}\\
f_\pi(\omega) &=& \left(1+({\omega'\omega}/{m^2})\right)^{1/2} 
[-({e^2}/{4 \pi m})Z^2 +       
\omega\omega'({\alpha} - {\beta}) 
+{\cal O}({\omega}^2{{\omega}'}^2)], \label{fpi}\\
g_\pi(\omega) &=& \sqrt{\omega\omega'}[
({e^2}/{8 \pi m^2})  
( {\kappa}^2 + 4Z
{\kappa} + 2Z^2)
+ \omega\omega'
{{\gamma_\pi}} + {\cal O}
({\omega}^2 {{\omega}'}^2)]  \label{gpi}
\end{eqnarray}
where $Z e$ is the electric charge of the nucleon ($e^2/4\pi=1/137.04$), 
$\kappa$ the anomalous magnetic
moment of the nucleon and $\omega'=\omega/(1+\frac{2\omega}{m})$.

In the relations for $f_0(\omega)$ and $f_\pi(\omega)$ the first
nucleon structure dependent coefficients are the photon-helicity non-flip 
$(\alpha+\beta)$ (forward polarizability) and photon-helicity 
flip $(\alpha-\beta)$ (backward polarizability) linear
combinations of the electromagnetic polarizabilities $\alpha$ and
$\beta$. In the relations for $g_0(\omega)$ and $g_\pi(\omega)$
the corresponding coefficients are the spin-polarizabilities
$\gamma_0$ and $\gamma_\pi$, respectively.

The appropriate tool for the prediction of  polarizabilities 
is to simultaneously apply
the forward-angle sum rule for $(\alpha+\beta)$ and the backward-angle sum
rule for $(\alpha-\beta)$. This leads to the following relations
\cite{schumacher07a,schumacher09} : 
\begin{eqnarray}
&&\alpha=\alpha^s+\alpha^t, \label{pol1}\\
&&\alpha^s=\frac{1}{2\pi^2}\int^\infty_{\omega_0}\left[A(\omega)\sigma(\omega,E1,M2,\cdots)+B(\omega)\sigma(\omega,M1,E2,\cdots)\right]\frac{d\omega}{\omega^2},\label{pol2}\\
&&\alpha^t=\frac12(\alpha-\beta)^t
\end{eqnarray}
and
\begin{eqnarray}
&&\beta=\beta^s+\beta^t,\label{pol3}\\
&&\beta^s=\frac{1}{2\pi^2}\int^\infty_{\omega_0}\left[A(\omega)
\sigma(\omega,M1,E2,\cdots)+B(\omega)\sigma(\omega,E1,M2,\cdots)\right]
\frac{d\omega}{\omega^2},\label{pol4}\\
&&\beta^t=-\frac12(\alpha-\beta)^t,
\end{eqnarray}
with
\begin{eqnarray}
&&\omega_0=m_\pi+\frac{m^2_\pi}{2 m},\label{pol5}\\
&&A(\omega)=\frac12 \left( 1+\sqrt{1+\frac{2\omega}{m}} \right),\label{pol6}\\
&&B(\omega)=\frac12 \left( 1-\sqrt{1+\frac{2\omega}{m}} \right),\label{pol7}\\
&&(\alpha-\beta)^t=\frac{g_{\sigma NN}{\cal M}(\sigma\to\gamma\gamma)}{2\pi m^2_\sigma}
+\frac{g_{f_0 NN}{\cal M}(f_0 \to\gamma\gamma)}{2\pi m^2_{f_0}}
+\frac{g_{a_0 NN}{\cal M}(a_0\to\gamma\gamma)}{2\pi m^2_{a_0}}\tau_3.
\label{pol8}
\end{eqnarray}
In (\ref{pol1}) to (\ref{pol8}) $\omega$ is the photon energy in the
lab system and  $m_\pi$
the pion mass.  The quantities $\alpha^s,\beta^s$
are the $s$-channel electric and magnetic polarizabilities, and 
$\alpha^t,\beta^t$ the $t$-channel electric and magnetic polarizabilities,
respectively. The multipole content of the photo-absorption cross section
enters through
\begin{eqnarray}
&&\sigma(\omega,E1,M2,\cdots)=\sigma(\omega,E1)+ \sigma(\omega,M2)+\cdots,
\label{pol9}\\
&&\sigma(\omega,M1,E2,\cdots)=\sigma(\omega,M1)+ \sigma(\omega,E2)+\cdots,
\label{pol20}
\end{eqnarray}
 i.e. through the sums of cross sections with change and without 
change of parity during the electromagnetic transition, respectively. The
multipoles belonging to parity change are favored for the
electric polarizability $\alpha^s$ whereas the multipoles belonging to parity
nonchange are favored for the magnetic polarizability $\beta^s$. For the
$t$-channel parts we use the pole representations described in
\cite{schumacher07b}. 

The forward spin-polarizability is given by
\begin{equation}
\gamma_0=-\frac{1}{4\,\pi^2}\int^\infty_{\omega_0}\left[\sigma_{3/2}(\omega)
-\sigma_{1/2}(\omega)\right]\frac{d\omega}{\omega^3},
\label{forward}
\end{equation}
and the backward spin-polarizability  by \cite{schumacher07b,lvov99}
\begin{eqnarray} 
&&\gamma_\pi=\int^\infty_{\omega_0}\sqrt{1+\frac{2\omega}{m}}\left(1+\frac{\omega}{m}
\right)\times \sum_n P_n[\sigma^n_{3/2}(\omega)-\sigma^n_{1/2}(\omega)]
\frac{d\omega}{4\pi^2\omega^3}+\gamma^t_\pi, \label{pol21}\\
&&\gamma^t_\pi=\frac{1}{2\pi m}\left[\frac{g_{\pi NN}
{\cal M}(\pi^0\to\gamma\gamma)}
{ m^2_{\pi^0}}\tau_3
+\frac{g_{\eta NN}{\cal M}(\eta \to\gamma\gamma)}{ m^2_{\eta}}
+\frac{g_{\eta' NN}{\cal M}(\eta'\to\gamma\gamma)}{
  m^2_{\eta'}}\right]\label{pol22}
\end{eqnarray}
where the parity factor is $P_n(E1,M2,\cdots)=-1$ and $P_n(M1,E2,\cdots)=+1$.
The quantities $g_{MNN}$ are the meson-nucleon coupling constants and 
${\cal M}(M\to\gamma\gamma)$ are the decay matrix elements.

\subsection{Introduction of electric and magnetic spin-polarizabilities}

It has become customary to describe the spin-independent polarizabilities 
via an electric ($\alpha$) and magnetic ($\beta$) polarizability 
and the spin-dependent polarizabilities via a forward spin-polarizability
$\gamma_0$ and a backward spin-polarizability $\gamma_\pi$. During the present
studies we noticed that it is also useful to separate the
spin-polarizabilities into an  electric and a magnetic part. One argument
in favor of the introduction of electric and magnetic spin-polarizabilities
may be derived from the fact that a large portion of the spin-polarizability 
is due to the $E_{0+}$  amplitude which does not have a transparent
relation to the spin of the nucleon. This  amplitude  is related 
to meson photoproduction via an electric-dipole excitation of the 
nucleon-pion system and, therefore, in a natural way demands the introduction
of a spin-polarizability $\gamma_E$. Related considerations may be found in
\cite{holstein00,hildebrandt04a,hildebrandt04b}.

\begin{figure}[h]
\begin{center}
\includegraphics[width=0.5\linewidth]{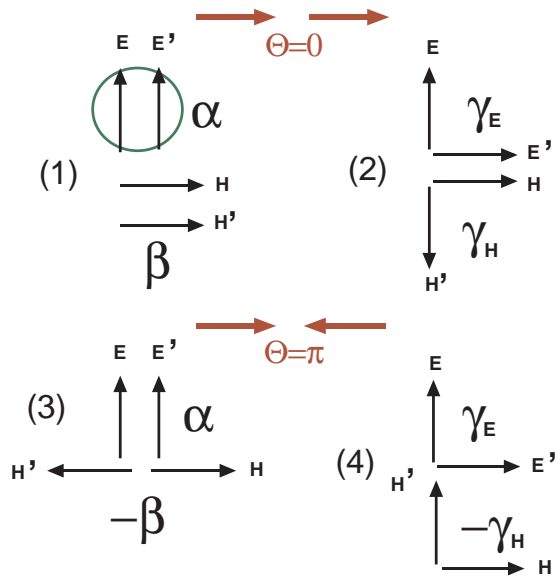}
\end{center}
\caption{Compton scattering viewed as simultaneous interaction of two electric
field vectors ${\bf E,E'}$ and magnetic field vectors ${\bf H,H'}$ for four
different cases. (1) Helicity independent forward Compton scattering as given
by the amplitude $f_0(\omega)$. (2) Helicity dependent forward Compton 
scattering as given by the amplitude $g_0(\omega)$. (3) Helicity independent
backward Compton scattering as given by the amplitude $f_\pi(\omega)$.
(4) Helicity dependent backward Compton scattering as given by the amplitude
$g_\pi(\omega)$. Longitudinal photons can only provide two electric vectors
with parallel planes of linear polarization as shown in the encircled  upper
part of panel (1). In panels (2) and (4) the direction of rotation leading
from ${\bf E}$ to ${\bf E'}$ depends on the helicity difference
$|\lambda_N-\lambda_\gamma|$, being 1/2 or 3/2. }
\label{ehfields}
\end{figure}

Polarizabilities may be measured by simultaneous interaction of two photons
with the nucleon. This is depicted in Figure \ref{ehfields}.
 The scattering
of slow neutrons in the electrostatic field of a heavy nucleus corresponds to
the encircled  upper part of panel (1) in Figure 1, showing two parallel
electric vectors. Only this case is accessible with longitudinal photons
at low particle velocities.
Compton scattering in the forward and backward directions leads
to more general combinations of electric and magnetic fields. 
This is depicted in the four panels of  Figure \ref{ehfields}.
Panel (1) corresponds to the amplitude $f_0(\omega)$, panel (2) to the
amplitude $g_0(\omega)$, panel (3) to the amplitude $f_\pi(\omega)$ and
panel (4) to the amplitude $g_\pi(\omega)$. Panel (1) contains two parallel
electric vectors and two parallel magnetic vectors. Through these
the electric polarizability $\alpha$ and the magnetic polarizability $\beta$ 
can be measured. In panel (3) corresponding to backward scattering the
direction of the magnetic vector is reversed so that instead of $+\beta$
the quantity $-\beta$ is measured. The nucleon  has a spin and because of 
this the
two electric and magnetic vectors have the option of being perpendicular
to each other.  This leads to the definition of the spin-polarizability
$\gamma$ which comes in different versions $\gamma_E$ and $\gamma_H$,
respectively. The different directions of the magnetic field in the forward
and backward direction leads to definitions of polarizabilities for the
four cases
\begin{eqnarray}
&&(1)\quad (\alpha+\beta)\label{eq11}\hspace{20mm}
\text{forward polarizability},
\label{forpol}\\
&&(2) \quad \gamma_0= \gamma_E+\gamma_H\hspace{9mm}
\text{forward spin-polarizability}, 
\label{forspin}\\
&&(3)\quad (\alpha-\beta)\hspace{20mm}
\text{backward polarizability}, \label{backpol}\\
&&(4) \quad \gamma_\pi=(\gamma_E - \gamma_H)\hspace{6mm}
\text{backward spin-polarizability}. \label{backspin}
\end{eqnarray}

\section{Resonant and nonresonant spin-dependent and spin-independent cross
  sections}

The $s$-channel degrees of freedom are given by the resonant and nonresonant
spin-dependent and spin-independent cross sections for photo-absorption.
The most precise information on these cross sections may be obtained from
the CGLN amplitudes of meson photoproduction and from the 
couplings $A_{1/2}$ and  $A_{3/2}$ of the nucleon resonant states. 
The somewhat lengthy procedure to obtain  these relations between partial
cross sections and the amplitudes is described in detail
in the previous paper \cite{schumacher09} and should not be repeated here. 
Instead we give in the following a detailed description of the essential
results and refer the reader to the previous work \cite{schumacher09}
for further details.

\subsection{Prediction of nonresonant photo-absorption cross section}

In case of the nonresonant contributions we have to take into account the
CGLN amplitudes of the  multipoles $E_{0+}$, $M^{(3/2)}_{1-}$ and 
$(M,E)^{(1/2)}_{1+}$. 
These amplitudes lead to the following partial cross sections
\begin{eqnarray}
&&\sigma_{1/2}(0+)=\frac{8\pi q}{k}\left[3\left|_{(p,n)}E^{(1/2)}_{0+}\right|^2
+\frac23 \left|E^{(3/2)}_{0+}\right|^2\right], \label{E1}\\
&&\sigma_{3/2}(0+)=0, \label{E2}\\
&&\sigma^{(3/2)}_{1/2}(1-)=\frac{8\pi q}{k}\frac23\left|M^{(3/2)}_{1-}\right
|^2, \label{E3}\\
&&\sigma^{(3/2)}_{3/2}(1-)=0, \label{E4}\\
&&\sigma^{(1/2)}_{1/2}(1+)=\frac{8\pi
  q}{k}\frac32\left|3_{(p,n)}E^{(1/2)}_{1+}
+_{(p,n)}M^{(1/2)}_{1+}\right|^2, \label{E5}\\
&&\sigma^{(1/2)}_{3/2}(1+)=\frac{8\pi
  q}{k}\frac92\left|_{(p,n)}E^{(1/2)}_{1+}
-_{(p,n)}M^{(1/2)}_{1+}\right|^2 \label{E6}
\end{eqnarray}
where $q$ and $k$ are the 3-momenta of the pion and the photon in the
c.m. system, respectively.
From these relations the spin-dependent and spin-independent
cross sections can be calculated using available CGLN amplitudes.
For the $E_{0+}$  amplitude we obtain the relation
\begin{equation}
\sigma_{3/2}(0+)-\sigma_{1/2}(0+)=-2\,\sigma_T(0+).
\label{enullplus}
\end{equation}
Eq. (\ref{enullplus}) shows that the spin-dependent and spin-independent
amplitudes corresponding to the $E_{0+}$  multipole are proportional to
each other so that the following discussion of $\sigma_T(0+)$ implies also
the  parallel discussion of  $\sigma_{3/2}(0+)-\sigma_{1/2}(0+)$.

\begin{figure}[ht]
\centering\includegraphics[width=0.7\linewidth]{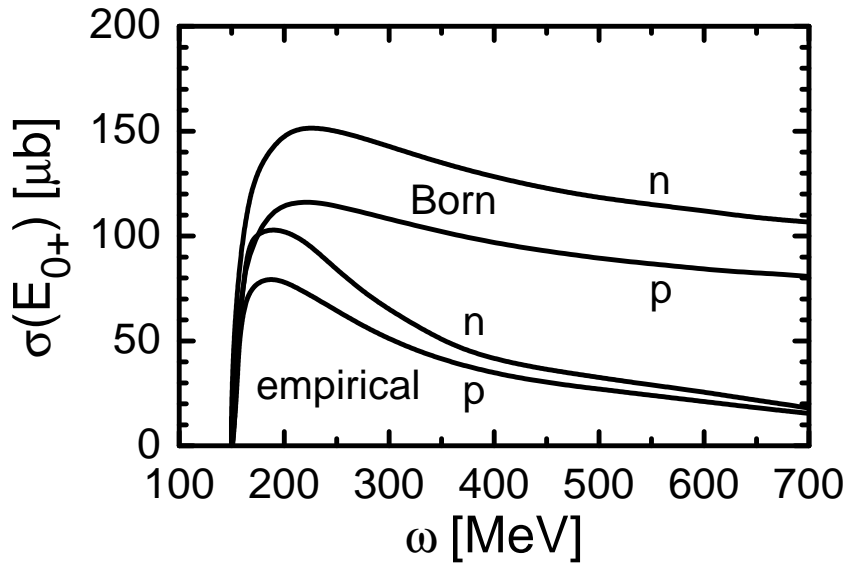}
\caption{Spin independent photo-absorption cross section due the $E_{0+}$
 amplitude for the neutron (n) and proton (p), respectively. 
Upper curves: Born approximation. Lower curves: Empirical 
data. }
\label{e0plus-5}
\end{figure}

The contribution of the $E_{0+}$  multipole may be considered as the
electric-dipole ``pion-cloud'' contribution because this is the only 
electric-dipole amplitude which is given
by nonresonant pion photoexcitation. 
In order to study the driving mechanism
for the photoproduction process it is useful to first refer to the Born 
approximation as represented by the two upper curves in Figure \ref{e0plus-5}.
In case of the Born approximation there  are two contributions 
{\it viz.} the Kroll-Ruderman term (photoproduction of charged pions 
on the nucleon) and the pion pole term (photoproduction 
of a charged pion being emitted by a nucleon). The difference between the
cross sections for the proton and the neutron is a consequence of the
different dipole moments of the $N\pi$ system in the two cases. 
For both contributions the pseudovector coupling (PV) constant
$f=g_{\pi NN}(m_\pi/2m)$ enters into the pion photoproduction matrix element
with $g_{\pi NN}=13.169\pm 0.057$
(for more details see \cite{ericson88}).

In Figure \ref{e0plus-5} the upper cross sections predicted in the Born 
approximation 
apparently are much larger than the empirical cross sections. This means
that the ``pion cloud'' contribution to the polarizabilities and 
spin-polarizabilities cannot simply be understood in terms of the
Kroll-Ruderman
term and the pion pole term together with the pseudovector coupling constant
$f$.  Two reasons for the deviation of the empirical $E_{0+}$ amplitude from
the Born approximation have been discussed in \cite{drechsel99}.
The first reason is that the pseudovector (PV) coupling is not valid at high
photon energies but has to be replaced by some average  of the PV and the
pseudoscalar (PS) coupling, or by introducing a formfactor. 
The second reason are $\rho$- and $\omega$-meson
$t$-channel exchanges which are not taken into account in the Born
approximation. From these findings we have to conclude that the ``pion
cloud'' contribution to the polarizabilities and spin-polarizabilities 
cannot be understood in terms of models which only take into account
those effects  which show up in the low-energy limit.

\subsection{Predicted  resonant photo-absorption cross sections}

For the resonant cross sections in principle also a direct calculation 
from the CGLN amplitudes is possible provided the one-pion branching 
$\Gamma_\pi/\Gamma_r$ of the resonances is taken into account where necessary.
However, as shown in  \cite{schumacher09}, a much more
precise procedure is available. This procedure makes use of the fact
that precise values for total cross-sections $\sigma_T$ are easier to obtain
than for the cross-section differences $\sigma_{3/2}-\sigma_{1/2}$. Therefore,
we start from the relation
\begin{equation}
\sigma^{n}_{3/2}-\sigma^{n}_{1/2}=A_n\,\frac12\,
(\sigma^n_{3/2}+\sigma^n_{1/2}) =A_n\,\sigma^n_T
\label{crossansatz}
\end{equation}
where $n$ refers to the different resonant states. For the precise prediction
of $\sigma^{n}_T =\frac12\,
(\sigma^n_{3/2}+\sigma^n_{1/2})$ we use the Walker parameterization of resonant
states where the cross sections are represented in terms of Lorentzians
in the form
 \begin{equation}
I=I_r\left(\frac{k_r}{k}\right)^2\frac{W^2_r\,\Gamma(q)\Gamma^*_\gamma(k)}
{(W^2-W^2_r)^2 +W^2_r\,\Gamma^2(q)}
\label{peakcross1}
\end{equation}
where $W_r$ is the resonance energy of the resonance. The functions
$\Gamma(q)$ and $\Gamma^*_\gamma(k)$
are chosen such that a precise representation of the shapes
of the resonances are obtained. The appropriate parameterizations and the 
related references are given in \cite{schumacher09}.
Furthermore, some consideration given in \cite{schumacher09}
shows that the peak cross section can be expressed through the resonance
couplings in the following form
\begin{equation}
I_r=\frac{2\,m}{W_r \Gamma_r}\left[|A_{1/2}|^2+|A_{3/2}|^2\right].
\label{peakcross2}
\end{equation}
For the total
widths $\Gamma_r$ and the resonance couplings 
precise data are available.
Tabulations of values adopted for the present purpose are given in
\cite{schumacher09}. 
\begin{figure}[ht]
\centering\includegraphics[width=0.47\linewidth]{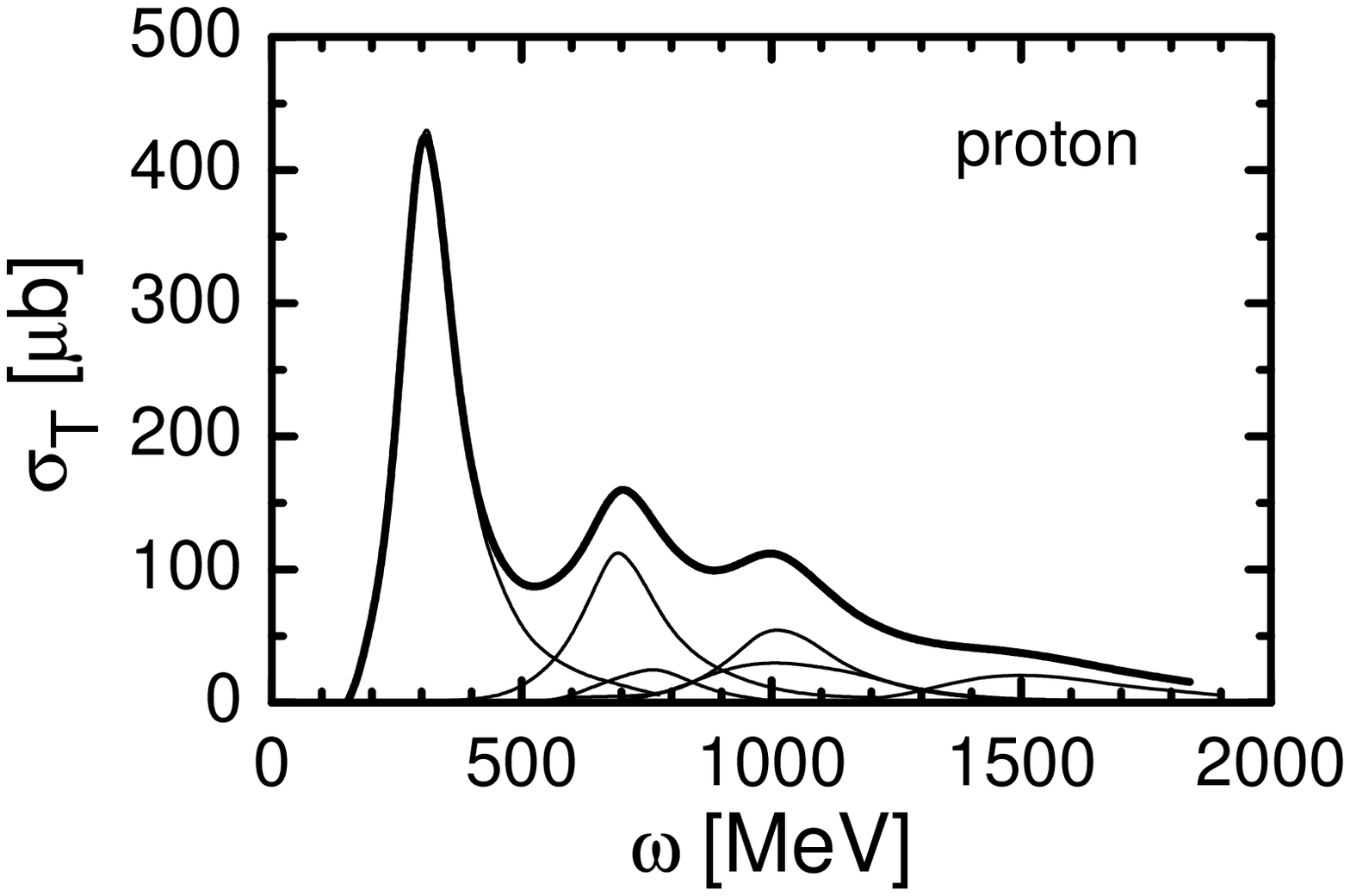}
\centering\includegraphics[width=0.47\linewidth]{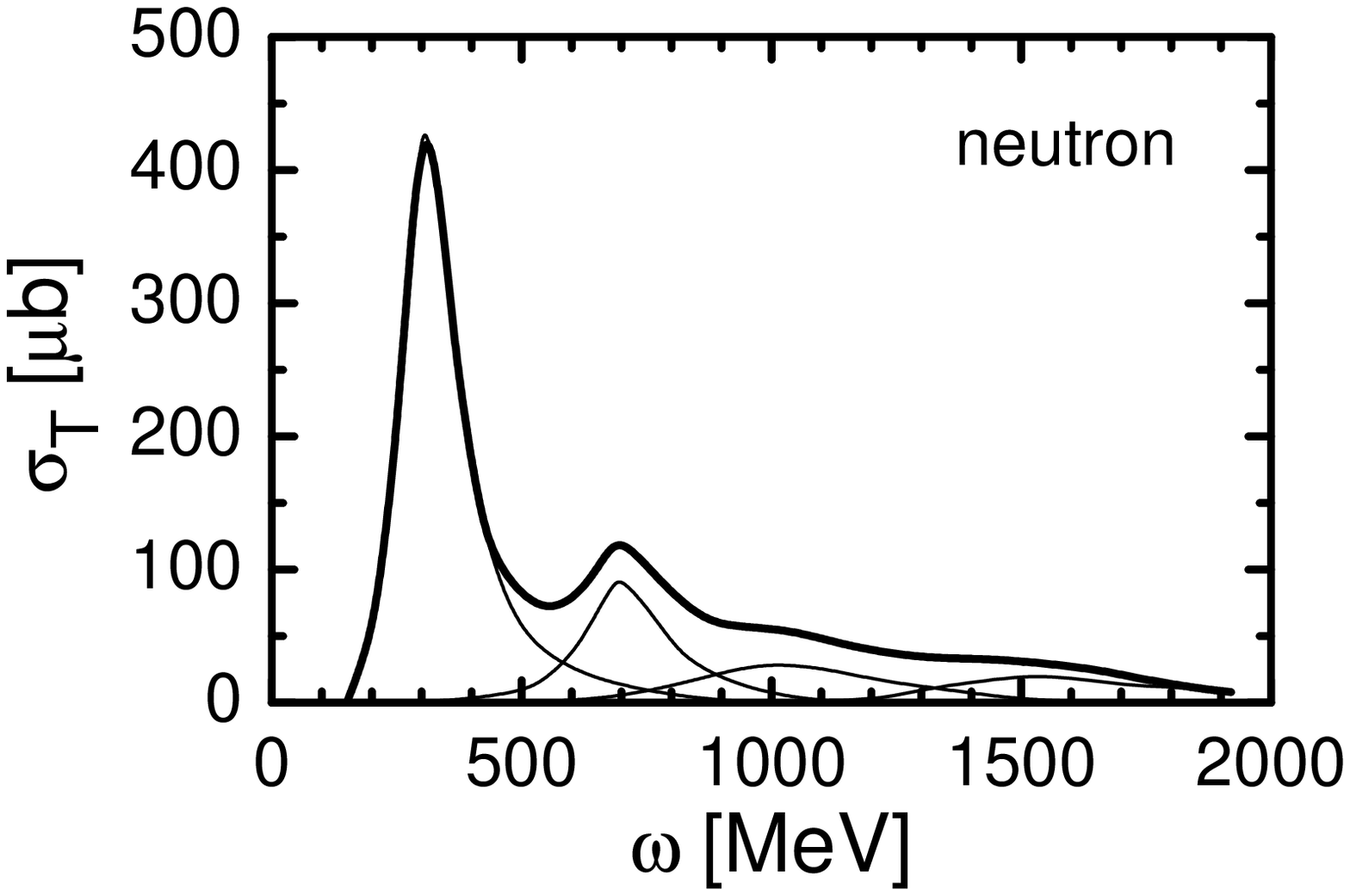}
\caption{Resonant part of the helicity independent
 photo-absorption cross-section for the
  proton and the neutron. Thick line: Sum of all resonances. Thin lines:
Contributing single resonances. The energies and strengths corresponding to
  these resonances are given in Tables 7 and 8 of \cite{schumacher09}.}
\label{coordinates1}
\end{figure} 
\begin{figure}[ht]
\centering\includegraphics[width=0.47\linewidth]{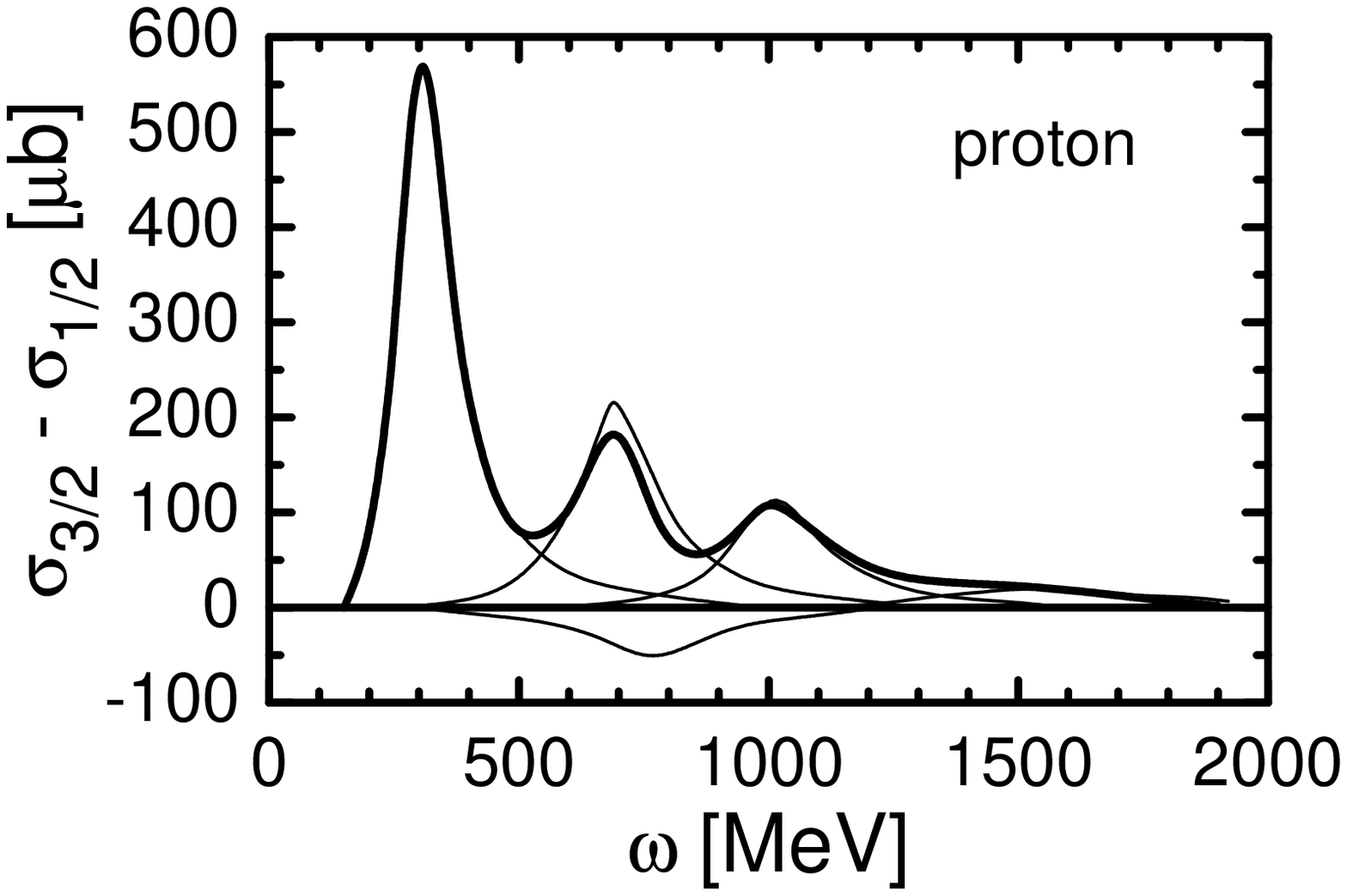}
\centering\includegraphics[width=0.47\linewidth]{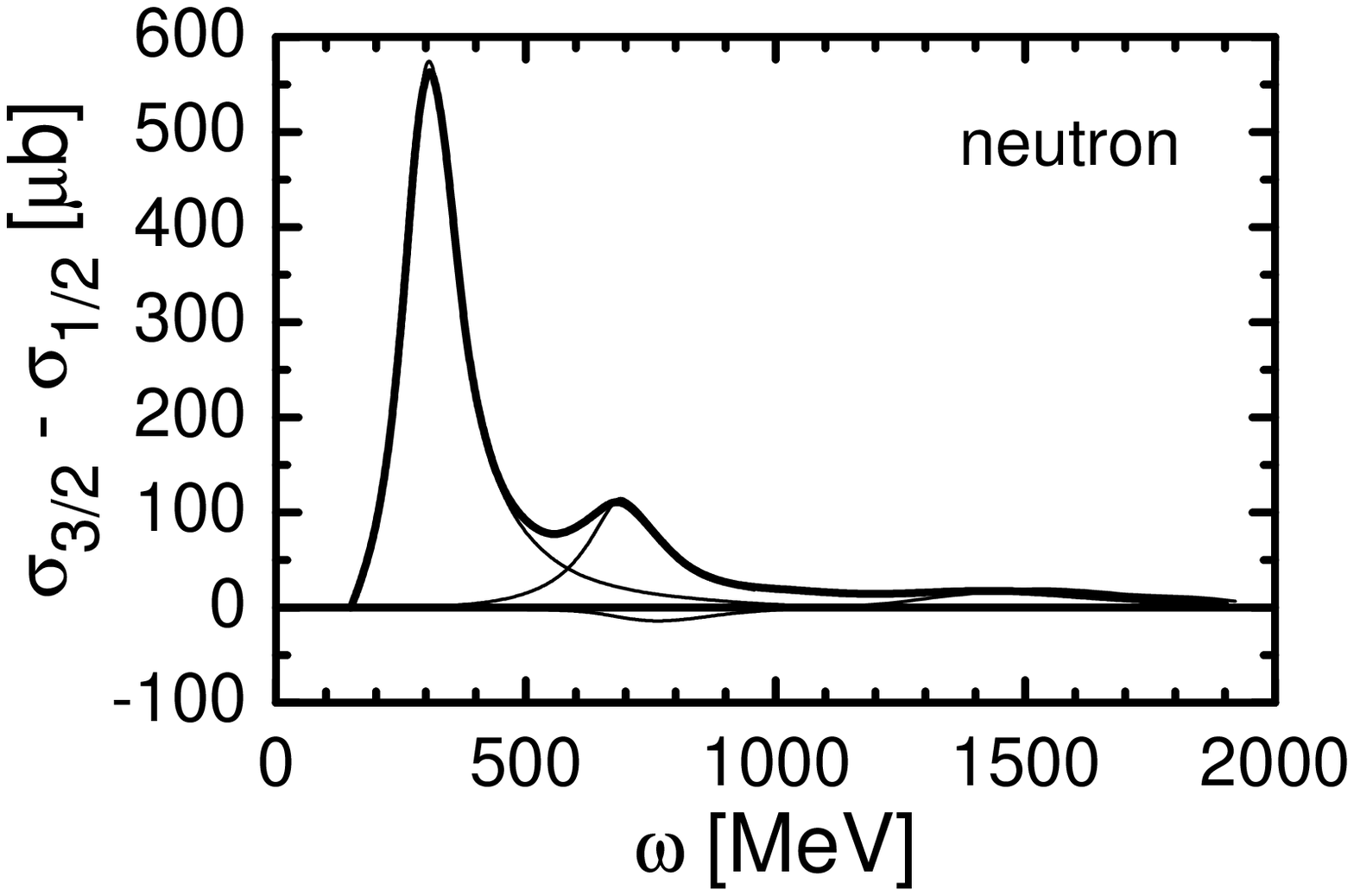}
\caption{Resonant part of the helicity dependent
 photo-absorption cross-section for the
  proton and the neutron. Thick line: Sum of all resonances. Thin lines:
Contributing single resonances. The energies and strengths corresponding to
  these resonances are given in Tables 7 and 8 of \cite{schumacher09}.
The signs of the cross sections are given in Table 3 of \cite{schumacher09}.
}
\label{coordinates2}
\end{figure} 
The results obtained for $\sigma_T$ and $\sigma_{3/2}-\sigma_{1/2}$ 
from the present procedure are shown in Figures \ref{coordinates1}
and \ref{coordinates2}.
The resonance couplings $A_{3/2}$ and $A_{1/2}$, the widths $\Gamma_r$
and the scaling factors $A_n$ entering into (\ref{crossansatz}) are tabulated
in \cite{schumacher09}. 
For the present purpose the  scaling factor $A_n$ of the $P_{33}(1232)$
resonance, i.e. quantity $A(P_{33}(1232))$
should be discussed in more detail because this resonance provides by far 
the largest contribution.

For this purpose we start from the relations
\begin{eqnarray}
&&\sigma^{(3/2)}_{1/2}(1+)=\frac{8\pi
  q}{k}\frac13 \left|3E^{(3/2)}_{1+}+M^{(3/2)}_{1+}\right|^2
  (\Gamma_r/\Gamma_\pi) 
\label{p33-1}\\
&&\sigma^{(3/2)}_{3/2}(1+)=\frac{8\pi
  q}{k}\left|E^{(3/2)}_{1+}-M^{(3/2)}_{1+}\right|^2 (\Gamma_r/\Gamma_\pi)
\label{933-2}
\end{eqnarray}
where the one-pion branching factor $(\Gamma_\pi/\Gamma_r)$ has been included
for completeness only, because this factor is equal to 1 in case of the
$ P_{33}(1232)$ resonance. The scaling factor $A_n$ 
(see Eq. \ref{crossansatz}) for the $P_{33}(1232)$ resonance may be obtained
in the
following way. First we introduce
\begin{equation}
\delta=\left(\frac{E^{(3/2)}_{1+}}{M^{(3/2)}_{1+}}\right)_{\rm res}
\label{emratio}
\end{equation}
and find
\begin{eqnarray}
&&\sigma^{(3/2)}_{1/2}(1+)=\frac12\,\sigma^{(3/2)}_T(1+)\,
(1+6\,{\rm Re}\,\delta),
\label{p33-3}\\
&&\sigma^{(3/2)}_{3/2}(1+)=\frac32\,\sigma^{(3/2)}_T(1+)\,
(1-2\,{\rm Re}\,\delta),
\label{p33-4}
\end{eqnarray}
\begin{equation}
A(P_{33}(1232))=1-6\,{\rm Re}\,\delta
\label{emratio-1}
\end{equation}
where use has been made of the reasonable 
approximation $|\delta|^2\ll |{\rm Re}\,\delta|$.
We see that $A(P_{33}(1232))=1$ for $\delta=0$,  but is strongly dependent
on $\delta$ otherwise. The number we get for $\delta$ is strongly 
dependent on the energy
at which we read $ {E^{(3/2)}_{1+}}$ and ${M^{(3/2)}_{1+}}$ from available
CGLN data. The appropriate choice is  to select that energy where 
the ratio $A(P_{33}(1232))$ obtains its maximum. Using the data in
\cite{drechsel07} we arrive at 
\begin{equation}
A(P_{33}(1232))=1.26 \quad \text{and} \quad {\rm Re}\,\delta=-4.4\%.
\label{numscal}
\end{equation}
This value for $A(P_{33}(1232))$ has been tested and found valid when compared
with the experimental Mainz data on $\sigma_{3/2}-\sigma_{1/2}$
\cite{schumacher09}. 
Therefore, it is
completely justified to use the numbers in Eq. (\ref{numscal}) for the
calculation of the spin-polarizabilities.

On the other hand the $E2/M1$ ratio given by 
${\rm Re}\,\delta$ is larger than the adopted value 
of  REM=$E2/M1$=$-2.2\%$ \cite{drechsel07} or 
REM=$-2.5\%$\cite{PDG}
by approximately a factor of 2. The explanation for
this difference is that REM is defined  at the resonance energy 
$W_r$ of the nucleon resonance where indeed the  smaller value for $E2/M1$
is obtained than the $E2/M1$ ratio corresponding to its maximum. 
For the comparison of the two different results for the $E2/M1$ ratio it is
useful to refer to \cite{tiator01} where it is clearly shown that at the
resonance energy $W_r= 1232$ MeV of the $P_{33}(1232)$ resonance
 Im $E_{1+}$ is equal to about $50\%$ of this quantity at its maximum
which is located at $W=1200$ MeV.

\section{Numerical results}

Numerical results for the spin-polarizabilities calculated on the basis
of the procedure given above and partly also outlined in \cite{schumacher09}
are given in Table  \ref{polresults2}.
\begin{table}[h]
\caption{Resonant (lines 2--9), single-pion nonresonant (lines 11--13)
and $t$-channel (lines 15--17)
components of the spin-polarizabilities in units of   $10^{-4}$fm$^4$.
For the backward spin-polarizabilities $\gamma_\pi$
more information may be found in \cite{schumacher09}. }
\begin{center}
\begin{tabular}{rl|rr|rr|rr|rr}
\hline
1&&$\gamma^{(p)}_0$&$\gamma^{(n)}_0$&$\gamma^{(p)}_\pi$&$\gamma^{(n)}_\pi
$&$ \gamma^{(p)}_E$&$\gamma^{(n)}_E$&$ \gamma^{(p)}_M$&$\gamma^{(n)}_M$\\ 
\hline
2&$P_{33}(1232)$&$-3.03$&   $-3.03$  &$+5.11$&$+5.11$&$+1.04$&$1.04$ 
&$-4.07$&$-4.07 $\\
3&$P_{11}(1440)$&$+0.05$&$+0.02$&$-0.10$&$-0.04$& $-0.025$&$-0.01$ 
&$+0.075$&$+0.03 $\\
4&$D_{13}(1520)$&$-0.14$&$-0.07$&$-0.39$&$-0.20$&$-0.265$&$-0.135$ 
&$+0.125$&$+0.65 $\\
5&$S_{11}(1535)$&$+0.05$&$+0.01$&$+0.13$&$+0.04$&$+0.09$&$0.025$ 
&$-0.04$&$-0.015 $\\
6&$S_{11}(1650)$&$+0.01$&$+0.00$&$+0.03$&$+0.00$&$+0.02$&$0.00$ 
&$-0.01$&$+0.00 $\\
7&$D_{15}(1675)$&$-0.00$&$-0.00$&$-0.00$&$-0.01$&$+0.00$&$-0.005$ 
&$0.00$&$+0.005 $\\
8&$F_{15}(1680)$&$-0.04$&$-0.00$&$+0.13$&$+0.01$&$+0.045$&$+0.005$
&$-0.085$&$-0.005 $\\
9& higher res.&$ 0.00$&$0.00$&$+0.03$&$+0.03$&$0.015$&$+0.015$ &$-0.015$
&$-0.015 $\\
\hline
10&sum-res.&$-3.09$&$-3.07$&$+4.94$&$+4.94$&$+0.925$&$+0.935$ 
&$-4.015$&$-4.005 $\\
\hline
11&$E_{0+}$&$+2.47$&$+3.18$&$+3.75$&$+4.81$&$+3.11$&$+3.995$ 
&$-0.64$&$-0.815 $\\
12&$M^{(3/2)}_{1-}$&$-0.11$&$-0.11$&$-0.18$&$-0.18$&$-0.145$&$-0.145$ 
&$+0.035$&$+0.035 $\\
13&$(M,E)^{(1/2)}_{1+}$&$+0.14$&$+0.38$&$+0.24$&$+0.66$&$+0.19$&$+0.52$ 
&$-0.05$&$-0.14 $\\
\hline
14&1$\pi$-nonres.&$+2.51$&$+3.45$&$+3.81$&$+5.29$&$+3.16$&$+4.37$ 
&$-0.65$&$-0.92 $\\
\hline
15&$\pi^0$-$t$-chan.&$0.00$&$0.00$&$-46.7$&$+46.7$&$-23.35$&$+23.35$ 
&$+23.35$&$-23.35 $\\
16&$\eta$-$t$-chan.&$0.00$&$0.00$&$+1.2$&$+1.2$&$+0.6$&$+0.6$ 
&$-0.6$&$-0.6 $\\
17&$\eta'$-$t$-chan.&$0.00$&$0.00$&$+0.4$&$+0.4$&$+0.2$&$+0.2$ 
&$-0.2$&$-0.2$ \\
\hline
18&sum $t$-chan.&0.00&0.00&$-45.1$&$+48.3$&$-22.55$&$+24.15$ 
&$+22.55$&$-24.15 $\\
\hline
19&$\gamma N\to \pi \Delta$&$0.00$&$0.00$&$-0.28$&$-0.23$&$-0.14$&$-0.115$ 
&$+0.14$&$0.115 $\\
\hline
20&tot. sum&$-0.58$&$+0.38$&$-36.6$&$+58.3$&$-18.6$&$+29.3$ 
&$+18.0$&$-29.0 $\\
\hline
\end{tabular}
\label{polresults2}
\end{center}
\end{table}
In this table we order the separate  contributions to the spin-polarizabilities
in three groups which are the resonant nucleon excitations, the nonresonant 
nucleon excitations and the $t$-channel contributions. The resonant
contributions are dominated by the $P_{33}(1232)$ resonance, the 
nonresonant nucleon excitations by the $E_{0+}$  amplitude and the
$t$-channel by the $\pi^0$ pole. 

Except for the precision, the advantage of the present method is that it is
comparatively easy to arrive at reliable errors. Three errors are relevant,
(i) the error of the spin-dependent peak cross section $1.26\times I_r$ 
of the $P_{33}(1232)$
resonance, the error of its  width $\Gamma_r$
and (ii) the  error of the cross section $\sigma(E_{0+}$).
These quantities are well  investigated, so that a 1$\sigma$ error 
of $3(4)\%$ for each of these 
quantities
 appears to be appropriate without being too optimistic. This leads to 
$\gamma^{(p)}_0=-0.58\pm 0.15(0.20)$ and $\gamma^{(n)}=+0.38\pm 0.17(0.22)$. 
Taking into account the well-known rules of error analysis these results are
not modified by the errors of all the other contributions even in case
rather large relative errors of the order of $20-30\%$ are adopted.
Therefore,
the values 
\begin{equation}
\gamma^{(p)}_0=-0.58\pm 0.20,\quad \gamma^{(n)}_0=+0.38\pm 0.22.
\label{res}
\end{equation} 
appear to be  justified as final results. These final results are also
given in the abstract.

For the proton the present result may be compared with the most recent
previous evaluation \cite{pasquini10}. The main result of this evaluation
is $\gamma^{(p)}_0=-0.90\pm 0.08 \pm 0.11$. Other results based on different 
photomeson analyses are $\gamma^{(p)}_0= -0.67$ (HDT), $-0.65$ (MAID), 
$-0.86$ (SAID) and $-0.76$ (DMT). In view of the fact that different 
data sets  have been used in these analyses the consistency of these results
and the agreement with our result
appears remarkably good. 
 A comparison is also possible
with the analysis of Drechsel et al. \cite{drechsel98} based on CGLN
amplitudes and dispersion theory. The numbers obtained are 
$\gamma^{(p)}_0=-0.6$ and $\gamma^{(n)}_0=+0.0$ based on the HDT
parametrization. The result obtained in
\cite{drechsel98} for the proton is in close agreement with our result,
not only with respect to the total result but also with respect to the 
partial contributions. The result obtained in \cite{drechsel98}
for the neutron confirms our result that the quantity $\gamma^{(n)}_0$
has the tendency of being shifted towards positive values.
In \cite{drechsel98}
also a detailed comparison with chiral perturbation theory is given which
should not be  repeated here.

Our present results for the backward spin-polarizabilities may be
compared with the predictions of L'vov and Nathan \cite{lvov99} obtained
on the basis of the SAID and HDT parameterizations of the CGLN amplitudes.
Our result agrees best with the results from the HDT parameterization
being $\gamma^{(p)}_\pi=-37.0$ and $\gamma^{(n)}=+57.8$. From these 
comparisons we conclude that the predicted 
spin-polarizabilities are satisfactorily 
known as far as their numerical values are concerned.

After obtaining this high precision for the  predicted backward
spin-polarizabilities 
a reconsideration of the corresponding experimental data is advisable.
 For the neutron the result $\gamma^{(n)}_\pi=58.6\pm 4.0$ given before
\cite{schumacher05} still remains its validity because there are no further
data available. For the proton three experiments of high precision
\cite{galler01,wolf01},\cite{olmos01} and  \cite{camen02}
have been carried out to determine $\gamma^{(p)}_\pi$. As mentioned above
the evaluation of the data requires the use of CGLN parameterizations to
represent the Compton amplitudes, in addition to the  spin-polarizability
$\gamma_\pi$ which is treated as an adjustable parameter. This procedure
implies that the determination of  $\gamma_\pi$
becomes to some extent model dependent. In our previous
determination of a {\it recommended} final result \cite{schumacher05}
all the available data have been included in the weighted average though some
of them showed large deviations from the majority of the data, 
which can be traced back to 
inconsistencies in the  respective CGLN
parameterizations. Details may be found in section 5.2 of Ref. 
\cite{schumacher05}. Figure 14 contained in the same section of Ref.
\cite{schumacher05} explains why results of an early 
experiment \cite{tonnison98} cannot be included in the averaging procedure.
Omitting now evaluations 
with obvious  inconsistencies the selection of data shown in Table
\ref{reevaluation} is obtained.
\begin{table}[h]
\begin{center}
\caption{Reevaluation of the experimental backward 
spin-polarizability of the proton:
The present average over experimental data differs from the previous one
\cite{schumacher05} because  some  evaluations of the data measured in  
\cite{galler01,wolf01} and \cite{camen02} 
have been  excluded because of systematic inconsistencies in the CGLN
amplitudes.}
\begin{tabular}{ll}
\hline
$\gamma^{(p)}_\pi$& reference\\
\hline
$-37.1\pm 3.0$& \cite{galler01,wolf01}\\
$-35.9\pm 2.3$& \cite{olmos01}\\
$-36.5\pm 2.5$&  \cite{camen02}\\
\hline
$-36.4\pm 1.5$ & weighted average\\
\hline
\end{tabular} 
\label{reevaluation}
\end{center}
\end{table}
We propose to use the weighted average shown in Table \ref{reevaluation}
as the updated  {\it recommended} value for the backward spin-polarizability
of the proton.

\section{Some properties of the polarizabilities and spin-polarizabilities}

The final goal of the ongoing  research is to eventually 
understand the polarizabilities and spin-polarizabilities in terms of 
models of the nucleon. We do not present a final solution for  this   
problem in the present paper. However, as a first step
we investigate some properties of the polarizabilities and 
spin-polarizabilities which may be helpful for  reaching the final goal.
A reasonable tool appears to us to compare some properties of 
polarizabilities and spin-polarizabilities with each other.

\subsection{Spin-polarizabilities compared with the electric and magnetic
  polarizabilities }

The introduction of electric ($\gamma_E$) and magnetic ($\gamma_M$)
spin-polarizabilities makes it possible to compare these quantities 
with the electric ($\alpha$) and magnetic ($\beta$) polarizabilities.
This is carried out in Tables 
\ref{electric} and \ref{magneticP33}.
In Table \ref{electric}  we investigate how the for quantities
$\alpha(E_{0+})$, $\beta(E_{0+})$, $\gamma_E(E_{0+})$
\begin{table}[h]
\caption{Electric and  magnetic polarizabilities
and spin-polarizabilities corresponding to the 
 excitation of the nonresonant $E_{0+}$ multipole.}
\begin{center}
\vspace{3mm}
\setlength{\extrarowheight}{10pt}
\begin{tabular}{l|ccc}
\hline
dispersion integral& proton & neutron&unit\\
\hline
$\alpha(E_{0+})=\frac{1}{4\pi^2}\int\left[1+\sqrt{1+\frac{2\omega}{m}}\right]
\sigma^{(E_{0+})}_T(\omega)\frac{d\omega}{\omega^2}$&+3.19
&+4.07&$10^{-4}$\,fm$^3$\\
$\beta(E_{0+})=\frac{1}{4\pi^2}\int\left[1-\sqrt{1+\frac{2\omega}{m}}\right]
\sigma^{(E_{0+})}_T(\omega)\frac{d\omega}{\omega^2}       $&--0.34&--0.43
&$10^{-4}$\,fm$^3$\\
$\gamma_E(E_{0+})=\frac{1}{4\pi^2}\int\left[1+\sqrt{1+\frac{2\omega}{m}}
(1+\frac{\omega}{m})\right]
\sigma^{(E_{0+})}_T(\omega)\frac{d\omega}{\omega^3}     
$&+3.11&+4.00&$10^{-4}$\,fm$^4$\\
$\gamma_M(E_{0+})=\frac{1}{4\pi^2}\int\left[1-\sqrt{1+\frac{2\omega}{m}}
(1+\frac{\omega}{m})\right]
\sigma^{(E_{0+})}_T(\omega)\frac{d\omega}{\omega^3} $
&--0.64&--0.82&$10^{-4}$\,fm$^4$\\
\hline
\end{tabular}
\end{center}
\label{electric}
\end{table}
\begin{table}[h]
\caption{Electric and  magnetic polarizabilities and spin-polarizabilities 
corresponding to
resonant excitation of the $P_{33}(1232)$ resonance.}
\begin{center}
\vspace{3mm}
\setlength{\extrarowheight}{10pt}
\begin{tabular}{l|ccc}
\hline
dispersion integral& proton & neutron&unit\\
\hline
$\alpha(P_{33})=\frac{1}{4\pi^2}\int\left[1-\sqrt{1+\frac{2\omega}{m}}\right]
\sigma^{(P_{33})}_T(\omega)\frac{d\omega}{\omega^2}$
&$-1.07$&$-1.07$&$10^{-4}$\,fm$^3$\\
$\beta(P_{33})=\frac{1}{4\pi^2}\int\left[1+\sqrt{1+\frac{2\omega}{m}}\right]
\sigma^{(P_{33})}_T(\omega)\frac{d\omega}{\omega^2}       $&$+8.32$&$+8.32$
&$10^{-4}$\,fm$^3$\\
$\gamma_E(P_{33})=-\frac{A(P_{33})}{8\pi^2}
\int\left[1-\sqrt{1+\frac{2\omega}{m}}
(1+\frac{\omega}{m})\right]
\sigma^{(P_{33})}_T(\omega)\frac{d\omega}{\omega^3}$
&$+1.04$&$+1.04$&$10^{-4}$\,fm$^4$\\
$\gamma_M(P_{33})=-\frac{A(P_{33})}{8\pi^2}
\int\left[1+\sqrt{1+\frac{2\omega}{m}}
(1+\frac{\omega}{m})\right]
\sigma^{(P_{33})}_T(\omega)\frac{d\omega}{\omega^3} $
&$-4.07$&$-4.07$&$10^{-4}$\,fm$^4$\\
\hline
\end{tabular}
\end{center}
\label{magneticP33}
\end{table}
and $\gamma_M(E_{0+})$
are related to the cross section $\sigma^{(E_{0+})}_T$. First of all we notice 
that dispersion integrals of very similar structure are obtained for the
polarizabilities and spin-polarizabilities. In the limit $m \to \infty$
only the electric parts $\alpha$ and $\gamma_E$ are different from zero.
This means that the nonzero values obtained for $\beta$ and $\gamma_M$ 
may be understood in terms of ``relativistic'' or ``recoil''
 effects. In a classical
model the quantity $\alpha$ corresponds to an electric dipole moment
induced in the ``pion cloud'' through the action of a first electric field
vector {\bf E}. This dipole moment interacts with an electric
field vector {\bf E}' being parallel to the direction 
of the first electric field vector      {\bf E}.             
The main difference between the electric polarizability $\alpha$ and the
electric spin-polarizability $\gamma_E$  is that for the latter 
quantity the second electric field vector {\bf E}'
is perpendicular to the first one. It certainly is a challenge for further
research to find a model which explains the relative sizes of the 
quantities $\alpha(E_{0+})$ and $\gamma_E(E_{0+})$. A model like this
may be expected to contain valuable information on the dynamics of the
excitation of the ``pion cloud''.

In Table \ref{magneticP33} it is investigated how the quantities 
$\alpha(P_{33})$, $\beta(P_{33})$, $\gamma_E(P_{33})$ and  $\gamma_M(P_{33})$
are related to the cross section $\sigma^{(P_{33})}_T$ corresponding to the
$P_{33}(1232)$ resonance. In this case the quantities $\beta(P_{33})$ and
$\gamma_M(P_{33})$ are the large quantities whereas the quantities 
$\alpha(P_{33})$ and $\gamma_E(P_{33})$ are small ``relativistic''
or ``recoil''  corrections.
The most interesting difference between the polarizabilities and the
spin-polarizabilities  is the enhancement factor $A(P_{33}(1232))$ 
which enters into
the spin-polarizabilities but does not enter into the polarizabilities. 
This enhancement factor has been discussed in detail in subsection 3.2. 
Since this factor differs from $A(P_{33}(1232))$=1 only because of the nonzero
$\delta=E^{(3/2)}_{1+}/M^{(3/2)}_{1+}$ ratio, it is of interest to study the 
quantity $\delta$ in terms of a model in order to eventually obtain a
deeper insight into the driving mechanisms connected with the
spin-polarizabilities. This is carried out in the next subsection.

\subsection{Experimental and predicted resonance couplings $A_{3/2}$ and
  $A_{1/2}$ for the $P_{33}(1232)$ resonance}

The resonance couplings are quantities which on the one hand can be determined
from experimental CGLN amplitudes and on the other hand can be predicted 
in models of the nucleon. Of these the SU(6)$\times$O(3)
harmonic oscillator model has been investigated in detail.
Therefore, the resonance couplings predicted in the framework of this model
are suitable
for  relating the resonant components of the spin-polarizabilities 
to a model of the nucleon. 

On the side of the experimental data the resonance couplings are given by
\begin{eqnarray}
&&A_{1/2}(1+)=-\frac12\left[3{\bar E}^{(3/2)}_{1+}+{\bar M}^{(3/2)}
_{1+}\right],
\label{recoup1}\\
&&A_{3/2}(1+)=\frac12\sqrt{3}
\left[{\bar E}^{(3/2)}_{1+}-{\bar M}^{(3/2)}_{1+}\right]
\label{recoup2}
\end{eqnarray}
where
\begin{equation}
({\bar E}^{(3/2)}_{1+},{\bar M}^{(3/2)}_{1+})=
\sqrt{\frac23}\left[\frac{4\pi\, q_r\, W_r\, \Gamma^2_r}{k_r\,m\,\Gamma_\pi}
\right]( E^{(3/2)}_{1+},M^{(3/2)}_{1+})
\label{recoup3}
\end{equation}
and where $q_r$ and $k_r$ are the 3-momenta $q$ and $k$ at the resonance 
maximum.
This leads to 
\begin{equation}
{\rm REM}=\frac{A_{1/2}(1+)-\frac{1}{\sqrt{3}}A_{3/2}(1+)}{A_{1/2}(1+)
+\sqrt{3}A_{3/2}(1+)}
=\frac{E^{(3/2)}_{1+}}{M^{(3/2)}_{1+}}.
\label{rem}
\end{equation}
Experimental data given in the literature may be found in Table
\ref{couplingtable-1}.
\begin{table}[h]
\caption{Experimental resonance couplings $A_{1/2}$ and $A_{3/2}$ 
in units of  $10^{-3}$GeV$^{-1/2}$ for the
$P_{33}(1232)$ resonance
taken from recent publications. The result
given by \cite{dugger07} has been adopted in \cite{schumacher09} and in the
present work to calculate the $\sigma_T(P_{33}(1232))$.}
\begin{center}
\begin{tabular}{cccl|c}
$A_{1/2}$  & $A_{3/2}$  &  $E2/M1$ $(\%)$ & Ref. 
&$E2/M1$ $(\%)$ Eq.(\ref{rem})\\
\hline
$ -135\pm 6$ & $ -250\pm 8$ & $-2.5\pm 0.5$ & \cite{PDG}&$-1.6$\\
$ -139.1\pm 3.6$ & $ -257\pm 4.6$ & $-$ & \cite{dugger07}&$-1.6$\\
$-140$& $-265$& $-2.2$& \cite{ drechsel07} &$-2.2$ \\
\hline
\end{tabular}
\end{center}
\label{couplingtable-1}
\end{table}
The $E2/M1$ ratios given in column  3 of Table \ref{couplingtable-1}
are the ones given by the
authors listed in column 4, whereas the $E2/M1$ ratios given in 
column  5 are calculated 
using Eq. (\ref{rem}). By comparing the results given in columns 3 and 5 
we see that there is only
partly consistency between these two types of data. This shows that our present
procedure to calculated $\sigma_T$ from the resonance couplings but not
$\sigma_{3/2}$ and $\sigma_{1/2}$ separately is justified. In 
\cite{schumacher09} we have adopted the results given by \cite{dugger07}
because of their precision and because these results  
exactly reproduce the value
$I_r(P_{33}(1232))=390$ $\mu$b directly  determined in a photo-absorption
experiment \cite{armstrong72}.

On the side of the theory the relevant helicity amplitudes for $N\to N^*$
transitions induced by the absorption of a real photon are (see e.g. 
\cite{thomas00})
\begin{eqnarray}
&& A_{1/2}=\langle N^*_{j,+\frac12}|H_{\rm int}|N_{\frac12,-\frac12}\rangle
\label{R1}\\
&& A_{3/2}=\langle N^*_{j,+\frac32}|H_{\rm int}|N_{\frac12,+\frac12}\rangle
\label{R2}
\end{eqnarray}
where $H_{\rm int}$ is the single-particle interaction Hamiltonian.
In a SU(6)$\times$O(3) quark model the resonance couplings for the 
$N(939)\to P_{33}(1232)$ are given by \cite{copley69}
\begin{eqnarray}
&&A_{1/2}=-\frac{2\sqrt{2}}{3}\mu\,\sqrt{\pi\,k}\,R^{s}_{00},\label{A1}\\ 
&&A_{3/2}=-2\sqrt{\frac{2}{3}}\mu\,\sqrt{\pi\,k}\,R^{s}_{00}\label{A2}
\end{eqnarray}
with
\begin{equation}
R^{s}_{00}=\exp{(-k^2/6\alpha^2)}
\label{wavefkt}
\end{equation}
where $\mu=0.13$ GeV$^{-1}$ is the quark scale magnetic moment in Gaussian
units, chosen to fit the magnetic moment of the proton and $\alpha^2=0.17$
GeV$^2$. The $g$ factor of the quarks is chosen to be $g=1$. Predicted results
are given in Table \ref{couplingstable-2}.
\begin{table}[h]
\caption{Predicted  resonance couplings $A_{1/2}$ and $A_{3/2}$ in
units of  $10^{-3}$GeV$^{-1/2}$
for the $P_{33}(1232)$ resonance
taken from previous work.}
\begin{center}
\begin{tabular}{lccl}
1&$A_{1/2}$  & $A_{3/2}$   & Ref. \\
\hline
2&$-101$&$-175$ & Copley et al. \cite{copley69} (nonrel.)\\
3&$-108$& $-187$& Feynman et al. \cite{feynman71} (rel.) \\
4&$-103$& $-178$& Feynman et al. \cite{feynman71} (nonrel.) \\
5&$ -113$ & $ -195$ &  Close, Li \cite{close90} (rel.)\\
6&$ -101$  & $-173$ & Close, Li \cite{close90} (nonrel.)\\
7&$ -108$ & $-186$ & Capstick \cite{capstick92} (rel.)\\
\hline
8&$-109.7$ & $ -189.3$ & average relativistic prediction\\
&&\\
9&$-139$& $-257$& exp. adopted \cite{schumacher09} \\
\hline
\end{tabular}
\end{center}
\label{couplingstable-2}
\end{table}
The results given in lines 2, 4 and 6 correspond to the formulae
(\ref{A1}) -- (\ref{wavefkt}) and are  named  nonrelativistic results
in the literature. The results 
in lines 3, 5 and 7 contain relativistic corrections of different kinds.
In the paper of Feynman et al. \cite{feynman71}  the nonrelativistic 
harmonic oscillator  Hamiltonian is replaced by a
relativistic or relativized version by introducing the 4-momenta of the quarks
instead of  3-momenta and by introducing coordinate operators. The results 
obtained
in this way for the resonance couplings are about $5\%$ larger than the
nonrelativistic results but still about $30-40\%$ smaller than the
experimental values. 
Similar findings have been made by Close and Li
\cite{close90} and by Capstick \cite{capstick92}.  
In the paper of Close and
Li \cite{close90} relativistic corrections due to spin-orbit
coupling are taken into account.
 In the paper of Capstick \cite{capstick92} an electromagnetic transition
 operator containing relativistic corrections is introduced in addition 
to relativized quark-model wave functions.
We see that the experimental data (line 9) are much larger 
than the predicted data
independent of the specific method of calculation. This means that
relativistic corrections cannot explain the large gap between the experimental
and theoretical results so that something else must be the reason.

Calculating the photon decay widths from the resonance couplings 
given in lines 8 and 9 of Table \ref{couplingstable-2}
we arrive at
\begin{equation}
\frac{\Gamma_\gamma({\rm theor.)}}{\Gamma_\gamma({\rm exp.)}}\simeq 0.56.
\label{widthP33}
\end{equation}
This means that the single particle model leads to a much too small photon
decay width independent of  the inclusion or omission of relativistic 
corrections
into the calculation. This discrepancy comes not as a surprise because the
quarks in the nucleon are strongly coupled to each other so that the single
particle transition  should be accompanied by a collective component. This
finding is a well-known phenomenon in nuclear physics. Furthermore, 
this picture
also allows to qualitatively explain the electric quadrupole ($E2$) component
of the transition which in a single particle model is difficult 
to understand.   
In connection with the $E2$ component our present study of the 
spin-polarizabilities has added an interesting further insight.  The $E2/M1$
ratio relevant for the enhancement of the cross-section difference 
$\sigma_{3/2}-\sigma_{1/2}$ is ${\rm Re}\,\delta=-0.044$ and thus a 
factor of 2 larger
that the standard value REM$=-0.022$. The reason for this difference is
that $\delta$ is defined to be the maximum possible $E2/M1$ ratio obtained from
the experimental CGLN amplitude whereas the  REM$=-0.022$ is determined
at the energy $W_r$, i.e. the resonance energy of the Lorentzian 
describing the 
nucleon resonance.

The results obtained in the foregoing may be summarized by making use of 
Eqs. (\ref{p33-3}) and (\ref{p33-4}). This leads to
\begin{eqnarray}
&&\sigma_{1/2}^{(3/2)}(1+)=\frac12\,\sigma^{(3/2)}_T(1+)\,
\Big(0.56(M1_{\rm sp})
+0.44(M1_{\rm coll})\Big)\Big(1-0.26(E2_{\rm coll})\Big), \label{res1}\\
&&\sigma_{3/2}^{(3/2)}(1+)=\frac32\,\sigma^{(3/2)}_T(1+)\,
\Big(0.56(M1_{\rm sp})
+0.44(M1_{\rm coll})\Big)\Big(1+0.09(E2_{\rm coll})\Big),\label{res2}\\
&&\sigma_{3/2}^{(3/2)}(1+)-\sigma_{1/2}^{(3/2)}(1+)=1.26\,\sigma^{(3/2)}_T(1+).
\label{res3}
\end{eqnarray} 
The cross sections  $\sigma_{1/2}^{(3/2)}(1+)$ and $ \sigma_{3/2}^{(3/2)}(1+)$
entering into
$\sigma_{3/2}^{(3/2)}(1+)-\sigma_{1/2}^{(3/2)}(1+)$ and, therefore, also into
the $P_{33}(1232)$ component of the spin-polarizabilities are composed
of single-particle (sp) parts and collective (coll) parts. The single-particle
parts have tentatively been identified with the prediction of the 
SU(6)$\times$O(3) harmonic oscillator model. The difference between 
this prediction
and the experimental total cross section $\sigma^{(3/2)}_T(1+)$ is tentatively
interpreted as being due to a collective  excitation mode which may be
understood as being due to a consequence of the strong coupling between 
the constituent quarks. This spin-dependent strong coupling 
also leads to a $E2$-component
of the electromagnetic transition which diminishes  the cross section
$\sigma_{1/2}^{(3/2)}(1+)$ by $26\%$ and increases the cross section
$\sigma_{3/2}^{(3/2)}(1+)$ by $9\%$. In total we arrive at the 
cross-section difference given in Eq. (\ref{res3}). The foregoing  discussion
may be compared with the work of Jenkins and Manohar \cite{jenkins94}
where relations among the baryon
magnetic and transition magnetic moments are derived in the $1/N_c$ expansion.
These results might be interpreted as
suggesting that the single-particle picture works quite well -- at least in
the case
of isovector $M1$ excitations of octet and decouplet baryons.
More insights into the dynamics of the $P_{33}(1232)$ resonance and 
of the $E_{0+}$ amplitude
may be
found in \cite{holstein00,hildebrandt04a,hildebrandt04b,pascalutsa07}.

\section{Summary and discussion}

In the foregoing paper we have predicted spin-polarizabilities for the proton
and the neutron and compared them with experimental data. Furthermore, we have 
related the spin-polarizabilities to excitation processes which for the 
$s$-channel is dominated by the nonresonant $E_{0+}$  amplitude and the 
 $P_{33}(1232)$ nucleon resonance. The $t$-channel contributions is dominated
 by  the $\pi^0$ pole contribution.    
It has been found to be very useful to introduce the
spin-polarizabilities $\gamma_E$ and $\gamma_M$, where the first quantity
can be attributed to a two-photon process with two perpendicular 
electric-field vectors and the second to a two-photon process 
with two perpendicular
magnetic-field  vectors. It is shown that the nonresonant $E_{0+}$  
amplitude 
makes a sizable contribution to $\gamma_E$, accompanied by 
a small relativistic 
correction to $\gamma_M$. For the  resonant contribution provided by the
$P_{33}(1232)$ resonance   the multipolarity dependence is opposite as
expected.
A deeper insight into the dynamics of the $E_{0+}$  and 
 $P_{33}(1232)$ nucleon resonance contributions of the spin-polarizabilities
 appear possible within models of the nucleon. In case of the 
$E_{0+}$  amplitude a model has to be found which relates the 
electric-dipole moment induced in the direction of the first electric vector
${\bf E}$ to a related electric dipole-moment in the direction of the
second  electric vector ${\bf E}'$ being perpendicular to ${\bf E}$. 
In case of the $P_{33}(1232)$ resonance
an  appropriate basis for a discussion  is provided by the  SU(6)$\times$O(3) 
harmonic
oscillator model. It is shown that this model predicts cross sections 
$\sigma_{1/2}^{(3/2)}(1+)$ and $\sigma_{3/2}^{(3/2)}(1+)$ which amount to only
$56\%$ of the $M1$ component of the corresponding experimental cross sections.
This observation leads in a natural way to the supposition that the strong
coupling between the constituent quarks  leads to a collective $M1$ component,
filling the gap of $44\%$ between single-particle prediction and the
$M1$ part of the experimental cross section. In addition to the effects
on the $M1$ excitation, the  spin-dependence of the 
strong coupling  between the constituent quarks leads 
to an $E2$ component, and through this to a diminishing
of the  $\sigma_{1/2}^{(3/2)}(1+)$ cross section by $26\%$ and to an increase
of the $\sigma_{3/2}^{(3/2)}(1+)$ cross-section by $9\%$. These effects
together lead to an increase of the cross-section difference                 
$\sigma_{3/2}^{(3/2)}(1+)-\sigma_{1/2}^{(3/2)}(1+)$ by $26\%$ 
in agreement with spin-dependent cross section measurements
carried out at MAMI (Mainz) \cite{ahrens01,schumacher09}. For the magnetic  
spin-polarizability $\gamma_M$
this means that the $E2$ excitation also leads to an increase by $26\%$.

\end{document}